\documentclass{article}
\usepackage[margin=1in]{geometry}

\usepackage{eucal}
\usepackage{algorithm}
\usepackage{algpseudocode}
\usepackage[round]{natbib}
\usepackage{xcolor}
\usepackage{amsmath}
\usepackage{amssymb}
\usepackage{graphicx}
\usepackage{siunitx,booktabs,authblk }
\usepackage{url}
\usepackage{hyperref}
\usepackage{setspace}

\AtBeginDocument{%
  }

\definecolor{colour1}{RGB}{166,206,227}
\definecolor{colour2}{RGB}{31,120,180}
\definecolor{colour3}{RGB}{178,55,250} 
\definecolor{colour4}{RGB}{51,160,44}

\newcounter{noteXXctr} 

\newcounter{noteZZctr} 

\newcounter{noteXLctr} 

\newcounter{noteMCctr} 

\newcommand{\cmid}{\,|\,}

\begin{document}

\title{Causality-Inspired Models for Financial Time Series Forecasting}


\author[1]{Daniel Cunha Oliveira \thanks{Corresponding author: doliveira@ime.usp.br}}
\author[2]{Yutong Lu}
\author[2]{Xi Lin}
\author[2]{Mihai Cucuringu}
\author[1,3]{André Fujita}
\affil[1]{University of São Paulo, Brazil}
\affil[2]{University of Oxford, UK}
\affil[3]{Kyushu University, Japan}

\date{\today}








\maketitle
\begin{abstract}
We introduce a novel framework to financial time series forecasting that leverages causality-inspired models to balance the trade-off between invariance to distributional changes and minimization of prediction errors. To the best of our knowledge, this is the first study to conduct a comprehensive comparative analysis among state-of-the-art causal discovery algorithms, benchmarked against non-causal feature selection techniques, in the application of forecasting asset returns. Empirical evaluations demonstrate the efficacy of our approach in yielding stable and accurate predictions, outperforming baseline models, particularly in tumultuous market conditions.

\end{abstract}

               \maketitle

\section{Introduction}
Financial forecasting is crucial for investment strategies, risk management, and economic planning. However, it is challenging due to the complexity, noise, and volatility of financial markets. Traditional forecasting models often fail to adapt to distributional shifts, leading to suboptimal decisions, especially in dynamic environments. These models typically prioritize minimizing prediction error on historical training data, but struggle with shifts in data distribution, a challenge exacerbated in financial time series due to the relatively small number of observations inherent in low-to-mid frequency models, i.e.,~when dealing with quarterly, monthly, or even daily data, on a large cross-section of assets. \citep{makridakis2018m4} have shown consistent difficulty for complex models to outperform simpler and more robust alternatives. 

Causality-inspired models offer a promising alternative by focusing on invariant features expected to hold across different market conditions. The work of \cite{peters2017} has emphasized the utility of causal reasoning to achieve robustness in such challenging conditions, inspiring a new generation of forecasting models that balance invariance to distributional changes with prediction accuracy.  Despite being a very promising approach in cross-sectional applications \citep{janzig2019,kyono2020,LopezPaz2016},   
their use in financial time series forecasting remains under-explored.  
Our work sheds light on the integration of causality within financial time series forecasting. This research leverages causal inference and invariant prediction techniques to improve asset returns forecasting. By focusing on invariant features, causality-inspired models promise greater robustness and reliability.



\textbf{Main contributions.} Our contributions are twofold. First, we propose a framework integrating causal inference with feature selection to enhance return predictions. Our models mitigate common issues like overfitting to historical data, spurious correlations, and regime shifts by focusing on invariant causal structures. Second, we compare state-of-the-art causal discovery algorithms with non-causal feature selection models in return forecasting, demonstrating that causally-driven models offer greater stability and outperform non-causal models, particularly during crises.

Our two-step approach begins with causal discovery to identify key predictors and culminates in constructing robust, invariant predictive models using these causal predictors. This method ensures that the models maintain predictive accuracy across various market conditions.

We conducted an empirical study using monthly data of SPDR S\&P 500 ETF (SPY) and macroeconomic indicators from 2000 to 2022. Our results show that causal discovery methods achieve comparable accuracy with fewer predictors. Notably, causal methods demonstrate remarkably low prediction errors during volatile periods, such as the global financial crisis (GFC) and the COVID-19 outbreak, showcasing their robustness in regime shifts. These causal models select more stable and meaningful sets of predictors throughout the sample period, indicating their reliability and consistency across varying market conditions. We also prove the economic benefits by implementing a strategy that trades SPY according to our model's predicted directions of market movements, leading to higher cumulative returns compared to the non-causal benchmark.

\textbf{Paper structure.} The remainder of this paper is organized as follows:. Section \ref{sec:related_work} reviews our contributions to the finance literature and causal discovery. Section \ref{sec:framework} introduces the proposed forecasting framework. Section \ref{sec:data} describes the data source. Section \ref{sec:experiment} explains our setup of empirical experiments. Section \ref{sec:results} presents the empirical results of return forecasting. Finally, Section \ref{sec:conclusion} summarizes the results and discusses limitations and future research directions.

\section{Related Work} \label{sec:related_work}

Forecasting asset returns is a continuously evolving field of active research with a long history, captivating the attention of both practitioners and academic researchers. The statistical properties of the return series \citep{tsay2009, cont2001empirical}, including low signal-to-noise ratios, heteroscedasticity, etc., make the forecasting tasks extremely challenging.

Traditional econometrics literature proposes various linear models, such as autoregressive integrated moving average (ARIMA) \citep{whittle1951hypothesis, ariyo2014stock}, vector autoregression (VAR) \citep{sims1980macroeconomics, stock2001vector}, etc. They incorporate lagged returns as well as external regressors for predictions. In the recent period, there has been a notable increase in the volume of literature exploring the utilization of non-linear machine learning methodologies \citep{henrique2019literature, kumbure2022machine, krauss2017deep} in the domain of return forecasting. \cite{gu2020} comprehensively study the performance of models, ranging from simple linear regression to neural networks, on predicting monthly equity returns using 94 company characteristics as features. 

A substantial body of literature has identified predictors exhibiting statistically significant predictability through rigorous in-sample testing. These include, but are not limited to, valuation ratios \citep{campbell1988dividend, rozeff1984dividend, kothari1997book, ang2007stock, campbell2006efficient, asness2000predicting}, historical returns and technical indicators \citep{lo1988stock, lo2000foundations, neftci1991naive, neely2014forecasting, lin2018technical, ccakmakli2016getting}, macroeconomic factors \citep{fama1977asset, li2002macroeconomic, flannery2002macroeconomic}, market microstructure indicators \citep{chordia2004order, chordia2002order, lu2024trade, lu2023co, cross_impact_OFI_QF}, and so forth. \cite{welch2008comprehensive} provide a comprehensive examination of the forecasting power of published predictors, emphasize the importance of out-of-sample tests, and suggest more variables to be tested for prediction. Subsequent works \citep{campbell2008predicting, lin2018technical, ccakmakli2016getting} reinforce existing variables and develop new predictors.

Despite the abundance of predictors, appropriate feature selection is critical when forecasting \citep{tsai2010combining, thawornwong2004adaptive} to avoid over-fitting, maintain robust out-of-sample performance, and improve interpretability. Popular techniques include regularized models such as LASSO (\citep{zhang2019forecasting, chinco2019sparse}), correlation-based methods (\citep{hall1999correlation}), etc. Causal methods offer a new perspective: predictions from a causal model generally remain accurate when the environment changes, whereas predictions from a non-causal model can be significantly off \citep{peters2016}. The invariance of causal models can potentially benefit financial forecasting in turbulent environments. Causal discovery, the process of identifying and inferring causal relationships from data, typically aims to recover the entire graph of causal relationships, including those not directly relevant to the outcome, focusing on false positive and false negative rates in identifying the relationships \cite{spirtes-zhang-2016}.  However, the earlier causal discovery methods  
did not emphasize using the causal graph for prediction until \cite{peters2016} and \cite{pfister2019} proposed an invariant causal prediction approach
. As \cite{de2023causal} suggests, understanding causation is essential in financial research, yet few studies have explored this \citep[e.g.]{zhang2014causal}. Causal inference holds great promise in identifying the proper drivers of asset returns from many predictors.

Suppose we intervene on the predictor variables or change the whole environment. The predictions from a causal model will in general work as well under interventions as for observational data. In contrast, predictions from a non-causal model can potentially be very wrong if we actively intervene on variables. The invariance of causal model can potentially benefit financial forecasting under turbulent environments. Causal discovery is the process of identifying and inferring causal relationships from data. Most of the existing causal discovery methods aim at recovering the entire graph encoding these causal relations among all variables and they aims the performance as a classification problem: they focus on the false positive and false negative rates of whether a causal discovery methods identifies a causal relationship. They did not really look into using the causal graph for prediction until \cite{peters2016} and \cite{pfister2019} proposed an invariant causal prediction approach to focus on using causal discovery for prediction. As \cite{de2023causal} suggests, causation is demanding in financial research. However, only a few explorations have been presented so far. Causal inference holds great promise in navigating the true drivers of the succeeding asset returns from the zoo of predictors. In addition, we attempt to propose a causal discovery method for feature selection in stock prediction.  

Our paper contributes to this field of research by examining state-of-the-art causal discovery and prediction methods in forecasting asset returns, identifying causal predictors beyond correlations from a large set of macroeconomic variables, and improving the accuracy and stability of forecasts.

\section{Forecasting Framework} \label{sec:framework}
Our forecasting framework comprises two steps: feature selection and forecasting. In this section, we start by describing the feature selection methods we use, and then we define the very simple model that uses the selected features in a prediction phase.

We first define notations for the following sections. Let $\mathcal{D}=\{ (Y_t, \boldsymbol{X}_t)_{t=1}^{T} \}$ be a time series dataset where $\boldsymbol{X}_t \in \mathbb{R}^{d}$ consists of $d$ features. Let $X_{i,t} \in \boldsymbol{X}_t$ be the $i$-th feature in the feature vector at time step $t$, and $\boldsymbol{X}^{-i}_t$ be the feature vector at time step $t$ excluding the $i$-th feature. Furthermore, let $\hat{Y}_t$ be the estimate or prediction of the target variable for a given timestep, and $\hat{Y}^{-i}_t$ be the estimate or prediction of the model without the $i$-th feature.

\subsection{Causal Feature Selection Algorithms}

This section will explain eight causal feature selection models that are part of our comparative study. We use a causal discovery method to identify predictors to be included in prediction models. We input raw features into the selected causal feature selection algorithm and output the subset of predictors with causal relations with the forecasting target.

A common characteristic among all the models we label as `causal' is their focus on identifying the direct causes of a specific target variable. This contrasts with feature selection models that rely solely on prediction error metrics for feature selection. One key property of feature selection models aimed at identifying the direct causes of a target is that the conditional distribution of the target given the direct causes remains unchanged when the distribution of all other variables, except for the target, is experimentally altered. We hypothesize that this property could lead to exciting robustness features for forecasting models utilizing these selected features \citep{peters2016}. 

Those models are at the core of the questions that this paper seeks to answer. The selection of these models aims to compare feature selection models on both sides of the prediction-invariance spectrum. The prediction side of the spectrum is a class of feature selection methods that seek to select features purely based on how feature inclusion/deletion would increase/decrease prediction error. On the other hand, the invariance side of the spectrum is a class of models that select features based on the direct relationship between these variables and the target variable. In the graphical causal model language, the set of features that characterize the class of methods in the invariance side of the spectrum are called causal parents, and, provided that a set of assumptions holds, they provide a shield against spurious correlations.

\subsubsection{Multivariate Granger Causality}

Multivariate Granger \cite{granger1969investigating} causality tests if a time series $X_{i,t}$ Granger-causes a time series $Y_t$ by comparing two models
\begin{align*}
    Y_t &= \sum_{j=1}^{p} B_j \boldsymbol{X}_{t-j} + \epsilon_t \\
    Y_t &= \sum_{j=1}^{p} B_j \boldsymbol{X}^{-i}_{t-j} + \epsilon^{'}_t
\end{align*}

The null hypothesis $H_0$ and alternative hypothesis $H_1$ are defined as
\begin{align*}
    H_0&: B_j = 0 \, \forall j \\
    H_1&: B_j \neq 0 \, \text{for some } j
    \end{align*}
Under $H_0$, the coefficients $B_j$ are zero, implying $X_{i,t}$ does not Granger-cause $Y_t$ \cite{lutkepohl2005}. The F-test is used to test the hypothesis, and the distribution under the null is an F-distribution.

\subsubsection{SeqICP}

Invariant causal prediction (ICP) \citep{peters2017} and ICP for sequential data (seqICP) \citep{pfister2019} originally inspired this work. It identifies causal relationships by finding predictors upon which the conditional distribution of $Y$ remains invariant across different environments or periods. The key assumption is causal invariance: there exists a set of variables $S^*$ such that the conditional distribution of the outcome given $S^*$ is consistent in all environments $e,h \in \mathcal{E}$
$$
Y^e\cmid \left(X_e^{S^*}=x\right) \stackrel{d}{=} Y_h\cmid\left(X_h^{S^*}=x\right) .
$$
This assumption prescribes no direct intervention on $Y$ in any environment. The method also assumes no unmeasured confounder affects the treatment-outcome relationship. SeqICP aims to estimate $S^*$, thus making invariant causal predictions resilient to environmental changes.

\subsubsection{VAR-LiNGAM}

VAR-LiNGAM \cite{hyvarinen2010estimation} combines vector autoregression with independent component analysis (ICA) to uncover causal relationships in time series data. For this model in particular, we will let $\boldsymbol{X}^{*}_t=( Y_t, \boldsymbol{X}_t )$, that is, the union of the target and the features. The model is defined as 
\begin{equation}
   \boldsymbol{X}^{*}_t = \sum_{j=0}^{p} A_i\boldsymbol{X}^{*}_t + \epsilon_t ,
\end{equation}
where $A_i$ is a matrix of coefficients, and $\epsilon_t$ are non-Gaussian independent error terms. The ICA estimates the model by separating the observed variables into maximally independent components. The algorithm involves the following steps:

\begin{enumerate}
    \item Estimating the mixing matrix $A_i$ using standard VAR for each lag.
    \item Applying ICA to the residuals to identify independent components.
    \item Inferring causal ordering from the independence of components.
    \item Re-estimating the VAR model using the inferred causal ordering for a final estimate of $A_i$.
\end{enumerate}

This approach leverages the temporal structure and non-Gaussianity of the data to infer causality.

Since VAR-LiNGAM requires the number of observations to be greater than the number of covariates, we employ clustering to include an extra dimension reduction step in empirical studies. The process involves conducting K-means clustering on the features, computing the sample Pearson correlation of the features with the target, and selecting the feature with the strongest correlation with returns within each cluster. This results in a single feature per cluster.

\subsubsection{Dynotears}

Dynotears combines static and dynamic Bayesian network structures. The model is formulated to capture temporal dependencies (dynamic component) and contemporaneous associations (static component). 

Let $\boldsymbol{X}^{*}_t=( Y_t, \boldsymbol{X}_t )$. Furthermore, we stack all $T$ observations of our time series the matrix $\boldsymbol{X}^* \in \mathbb{R}^{T \times d}$, and let $\boldsymbol{X}^{*}_{\text{lag}} = [ \boldsymbol{X}^{*}_1 | \cdots | \boldsymbol{X}^{*}_p ] \in \mathbb{R}^{T \times pd}$ be a matrix of lagged time series. The optimization procedure in Dynotears involves solving the following problem
\begin{align*}
    \min_{S, W} \quad & \frac{1}{2T}||\boldsymbol{X}^* - \boldsymbol{X}^*\boldsymbol{S} - \boldsymbol{X}^{*}_{\text{lag}}\boldsymbol{W}||_F^2 + \lambda(\|\boldsymbol{S}\|_1 + \|\boldsymbol{W}\|_1) \\
    \text{subject to} \quad & \boldsymbol{S} \text{ is acyclic},
\end{align*}
where $\boldsymbol{S} \in \mathbb{R^{ d \times d}}$ is the matrix representing static relationships, $\boldsymbol{W} = [ \boldsymbol{W}^{\top}_1 | \cdots | \boldsymbol{W}^{\top}_p ] \in \mathbb{R}^{dp \times d}$ is the matrix representing dynamic relationships, $||\cdot||_F$ denotes the Frobenius norm, $||\cdot||_1$ denotes the L1 norm, and $\lambda$ is a regularization parameter that controls the sparsity of the matrices. The acyclic constraint on $S$ ensures that the learned structure is a Directed Acyclic Graph (DAG). The objective is to minimize the reconstruction error of $\boldsymbol{X}^*$ while promoting sparsity in the learned structures.

\subsubsection{PCMCI}
PCMCI \citep{runge2019detecting} is a constraint-based causal discovery algorithm that consists of two stages:
\begin{enumerate}
    \item PC$_1$, a robust variant of the PC algorithm \citep{spirtes1991algorithm,colombo2014order} adapted for time series, identifies an initial causal structure by iteratively removing edges based on conditional independence tests and orienting them using temporal constraints.
    \item Momentary Conditional Independence (MCI) tests refine this structure by validating or rejecting causal links at specific time lags, ensuring the relationships reflect actual temporal dependencies. This accounts for auto-correlation and controls false-positive rates. The significance of each link is assessed using P-values from the MCI test, which can be adjusted using procedures like false discovery rate control.
\end{enumerate}

\subsection{Traditional Feature Selection Algorithms}

In this section, we aim to explain the primary traditional feature selection model that will serve as a benchmark for the non-causal feature selection algorithm.

The key characteristic of this model is its purely prediction-based feature selection algorithm. In graph theory terminology, this feature selection model does not consider the direction of the edges from the features to the target. Consequently, we do not necessarily have the same robustness guarantees from models that seek invariant features (direct causes).

\subsubsection{Sequential Feature Selection}

Sequential Feature Selection (SFS) is a non-causal feature selection method that is our benchmark. It is a technique used in machine learning to select a subset of relevant features for model construction. It operates under the framework of a stepwise search to either incrementally add or remove predictors from the model and assess the impact on the model's performance \cite{whitney-1971}.

In the context of SFS, hypotheses are formulated to statistically test the significance of including/deleting a specific feature \( X_{i,t} \) by examining changes in a predefined loss function. 

Let $L(Y_t, \hat{Y}_t)$ be the chosen loss function. Then, the null hypothesis (\( H_0 \)) and the alternative hypothesis (\( H_1 \)) regarding the feature \( X_{i, t} \) are defined as follows
\begin{align*}
H_0 &: L({Y}_t, \hat{Y}_t) \leq L(Y_t, \hat{Y}^{-i}_t) \\
H_1 &: L({Y}_t, \hat{Y}_t) > L(Y_t, \hat{Y}^{-i}_t).
\end{align*}

Rejecting the null hypothesis means that we find evidence in favor of feature $i$ being able to reduce the loss, which leads to the inclusion of feature $i$ into the final model.

The models compared in sequential feature selection are generally of the form
\begin{align*}
\text{Model without } X_{i,t} &: Y_t = f(\boldsymbol{X}^{-i}_t) + \epsilon \\
\text{Model with } X_{i,t} &: Y_t = f(\boldsymbol{X}_t) + \epsilon'.
\end{align*}

The function $f$ can be set appropriately, and the SFS algorithm is independent of it. In this paper, we use a linear regression model and a random forest model as the base model of the SFS method.

Sequential feature selection can either be forward or backward. In the forward implementation, the process begins with an empty model and adds features iteratively. At each step, it adds the feature that improves the model the most until an addition does not improve it by a significant margin. In the backward implementation, the process starts with the full set of features and removes the least significant feature iteratively, which has the least effect on the model's performance.

We select the loss function as a pure prediction error loss, specifically the mean-squared error. This choice aims to differentiate between prediction-biased feature selection methods and invariance-based feature selection methods as much as possible.






\subsection{Forecasting Model}

With the selected predictors, $\mathbf{S}_{t} \in \mathbb{R}^{r}$, where $r < d$, we estimate a function, $g(\cdot)$, to forecast the target returns, $Y_{t+1}$,
\begin{equation*}
    Y_{t+1} = g (\mathbf{S}_{t}) + \epsilon_{t+1},
\end{equation*}
where $\epsilon_{t+1}$ are the residuals. Generally, $f(\cdot)$ can be any forecasting model. As we focus more on the performance of causal future selection, we adhere to linear models in this study, that is
\begin{equation*}
    Y_{t+1} = \mathbf{S}_{t} \boldsymbol{\beta} + \epsilon_{t+1},
\end{equation*}
where we estimate the regression coefficients, $\boldsymbol{\beta}$, using the ordinary least square (OLS). 

\section{Data Description} \label{sec:data}

Our empirical study is based upon SPY returns and macroeconomic indicators at the monthly level spanning the period from 2000-01 to 2022-12. We extract price data from the WRDS and macroeconomic data from the Federal Reserve Economic Data (FRED) databases. This section provides a detailed description of the data source and the preprocessing steps.

\subsection{SPY Data} 

The dataset we are looking at includes daily trading data for the SPDR S$\&$P 500 ETF (SPY). The SPY tracks the S$\&$P 500 index, providing a comprehensive overview of the US stock market. The dataset comprises daily observations, offering insights into both short-term fluctuations and long-term trends within these sectors.

This dataset contains monthly closing prices of SPY from February 2nd, 2000, to April 28th, 2023, and is collected from Bloomberg. 

\subsection{Macroeconomic Data}

The FRED-MD dataset \cite{mccracken2016} is a comprehensive collection of macroeconomic data curated by the Federal Reserve Bank of St. Louis. It is part of the broader FRED database, specifically supporting macroeconomic research and analysis. The dataset includes a wide array of time series data covering various aspects of the U.S. economy.

FRED-MD includes a wide range of economic indicators, such as: (1) Output and Income, (2) Consumption, Orders, and Inventories, (3) Labor Market, (4) Housing, (5) Money and Credit, (7) Interest and Exchange Rates, (8) Prices. We exclude group (6) because we already include lags of the ETF returns in the feature set.

The dataset comprises a time series with monthly observations, providing a detailed and nuanced view of the U.S. economy. The time frame spans several decades, offering a longitudinal perspective on economic trends and cycles. The dataset begins on January 1st, 1960, and concludes on December 30th, 2022.

The FRED-MD database also accompanies metadata that, among other things, suggests proper transformations to be applied to the time series, provides a brief description, and indicates a group to which the series belongs.
\section{Experimental Setup} \label{sec:experiment}
This section describes the data pre-processing for the ETF and the macroeconomic features. Furthermore, we present the main performance evaluation measures appropriate to answering the main questions of this work. Finally, we carefully describe each step of the training procedure, which involves (1) feature filtering (when needed), (2) feature selection algorithms, and (3) forecasting, given the selected features.
\subsection{Data Pre-processing}
Our objective during the data processing stage is to combine the daily price data of SPY with the monthly FRED-MD macro data. It is worth noting that the FRED-MD dataset has already arranged all the data points with vintages and the right release dates. Therefore, our main challenge is to apply the necessary transformations to the FRED-MD data and merge it with the SPY.

The process involves the following steps. The SPY prices data are available on the last business day of every month, whereas the FRED-MD data is on the first business. Furthermore, the dates in the FRED-MD data represent the data collection period, so where one sees, for instance, an actual value on January 1st, 2022, it means that this actual value was collected in January 2022 but released in February 2022. Thus, to avoid look-ahead bias, we shift the FRED-MD data one month forward. In practice, this means that for every month, we shift the available information from the first calendar day to the last business day. After doing that, the macroeconomic data is ready to be merged with the SPY returns.

\subsection{Training Procedure}

We repeat the forecasting procedures in Section \ref{sec:framework} for each month of the training sample until the very last month. We use a training window $w$ and incrementally increase it after the prediction phase in the last step. The sample's start is held fixed, thus characterizing a rolling window training procedure with a fixed start.

The first step of the training procedure entails using the raw features or the previously selected features derived from clustering to apply feature selection algorithms. These datasets are expanded to include the first lag of the current ETF return and the p-th lags of the macroeconomic features. The augmented dataset is then fed through the feature selection algorithms, which output the selected feature names.

The second step involves utilizing the previous step's selected features for prediction and the trading strategy. We estimate a linear regression model using ordinary least squares to make predictions. After making predictions for the next months, we store the predictions, increase the training window $w+1$ by one month, and repeat the process.

\subsection{Performance Evaluation}

In this research, we address the questions using two sets of evaluation metrics commonly found in financial machine-learning literature: statistical and economic performance evaluation.

For statistical performance evaluation, we utilize traditional forecasting error metrics typically employed in the context of regression. These metrics include the root-mean-squared error (RMSE) and the mean absolute error (MAE). While these metrics are insightful, they do not address our main questions directly. We have hypothesized the existence of a trade-off between prediction error-based and invariance-based feature selection models. We employ a rolling version of the RMSE and MAE to explore this. The rolling RMSE for a given window size $h$ can be defined as
\begin{equation}
    L_{\text{Rolling RMSE}}(Y_t, \hat{Y}_t) = \sqrt{\frac{1}{h} \sum_{t=1}^{h} (Y_t - \hat{Y}_t)^2}
\end{equation}
and the rolling MAE can be defined similarly. This forecasting error metric enables us to check for performance deterioration when a dataset shift occurs, thus providing evidence in favor of or against the trade-off.

For the economic performance evaluation, we will develop a trading strategy based on the forecasts generated by each model and assess its performance using traditional portfolio metrics.

We intend to create a monthly rebalanced single ETF strategy based on the rolling-window forecasts. At the end of each month, we will take a long position on the ETF if the predicted return is positive and a short position if it is negative. The return on the strategy is calculated as
\begin{equation*}
    R_{t} = sgn(\widehat{Y}_{t}) Y_{t},
\end{equation*}
where $sgn (\cdot)$ denotes the sign function. Upon computing the strategy returns, we will evaluate them using expected returns ($E[Rt]$), Sharpe ratio (Sharpe), and the Sortino ratio (Sortino) as long as the visual analysis of the time series of cumulative returns of equal-risk portfolios. We select these portfolio metrics, particularly because we will evaluate these during crisis and noncrisis periods; we avoid path-dependent measures such as the maximum drawdown.

\section{Empirical Results} \label{sec:results}

This section presents the experimental results, emphasizing the differences between unstable market and economic conditions and between crisis and regular periods. Notable crises include prolonged periods of the 2007--2008 Global Financial Crisis and COVID-19 pandemic over 2020--2021, as well as transient shocks such as the Tōhoku earthquake and tsunami in Japan in March 2011, Black Monday in August 2011, Chinese Black Monday in August 2015 and the Dow Jones plunge over February and March 2018. We use the approximate dates found by \cite{sulem-etal-2024} in a data-driven way as our crisis definition. Sequential feature selection with the Linear base model (SFS--Linear) is the benchmark non-causal feature selection method for comparison.

\subsection{Prediction accuracy}

\begin{figure}
    \centering
    \includegraphics[width=1\linewidth]{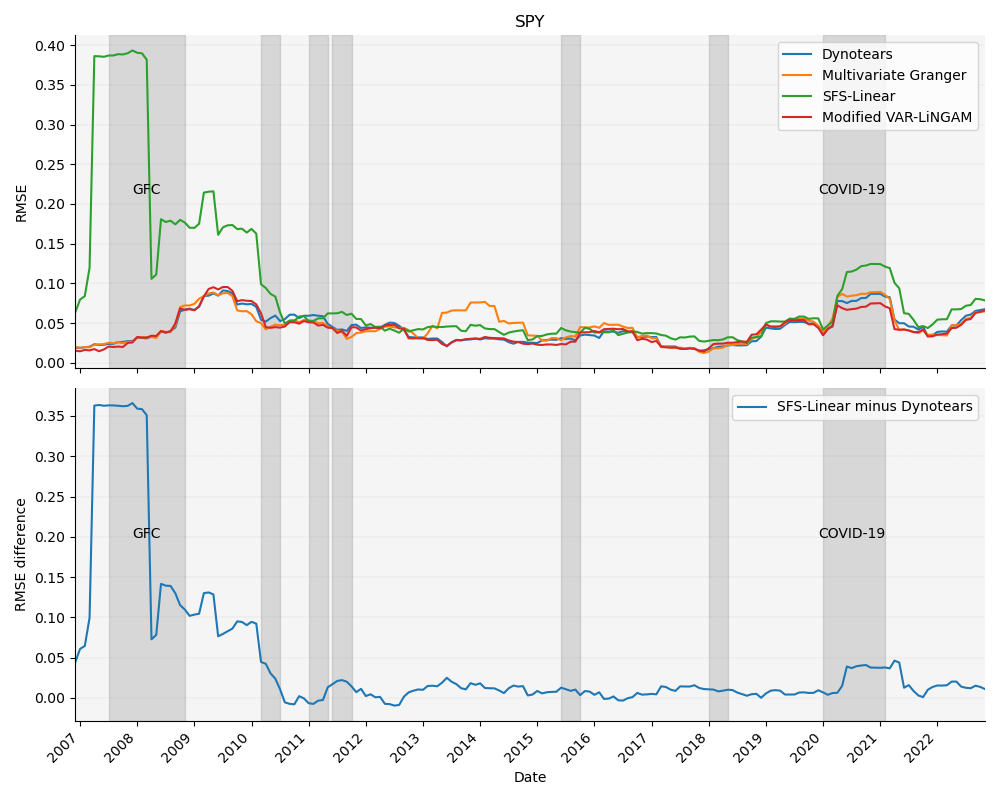}
    \caption{Top: Rolling RMSE of various SPY return prediction methods; Bottom: RMSE difference between Sequential Feature Selection (SFS-Linear) and Dynotears. The shaded areas indicate crisis periods.}
    \label{fig:mae_spy}
\end{figure}

The upper plot in Figure \ref{fig:mae_spy} compares the rolling RMSE of various methods over time, with shaded indicating crisis periods. Over the GFC and COVID-19 pandemic, the SFS--linear method consistently shows higher RMSE than causal methods. This is because crisis periods create different environments where the distribution of macroeconomic variables is affected by events, causing features merely correlated with returns to lose their predictive power or shift in relevance. In contrast, causal discovery algorithms identify variables that directly impact returns beyond mere correlation, maintaining invariant predictions even when distributions change.

During regular periods with stable economic conditions, the prediction accuracy of causal methods is similar to the benchmark. In stable environments, utilizing all associated variables for prediction is efficient and relatively safe, even if they do not directly cause movements in SPY returns.

Among the causal methods, Modified VAR-LiNGAM (red line) and the Dynotears (blue line) stand out among the models. The bottom plot in Figure \ref{fig:mae_spy} highlights the difference in RMSE between SFS-Linear and Dynotears. Values above 0 indicate that Dynotears yields a lower RMSE than SFS-Linear and vice versa. This comparison emphasizes that while causal and non-causal methods perform similarly under stable conditions, the advantage of invariant prediction becomes pronounced during crises, demonstrating the superiority of causal feature selection.

Table \ref{tab:model_metrics} summarizes the prediction metrics for regular periods and crises. There are some exciting findings to highlight. First, the Modified VAR--LiNGAM consistently outperforms (lowest prediction error) other methods in terms of MAE and RMSE, regardless of the period. As VAR--LiNGAM is the only method that includes a feature filtering step before application, it seems fair to showcase the performance of the other models excluding VAR-LiNGAM$^{*}$. Among these models, Dynotear stands out with the lowest prediction error metrics in almost all the analyzed periods, except for the MAE during crises, where Multivariate Granger (M--Granger) performs the best. It is worth noting that despite the overall superiority of the Modified VAR--LiNGAM$^{*}$, its results are very close to those of Dynotears. This implies that most of the performance difference between these models and SFS--Lin is influenced by the causal models' invariance rather than the filtering mechanism used in VAR-LiNGAM$^{*}$.

The last column of Table \ref{tab:model_metrics} depicts the MAE increase in percentage terms from regular periods to crisis periods. This metric can be understood as measuring the generalization error of those models on test data in crisis. It is possible to see that the prediction-based feature selection model, SFS-Lin, has the lowest generalization capabilities during crisis periods.

{\footnotesize
\begin{table}[h]
    \centering
    \begin{tabular}{@{}lSSSSS@{}}
        \toprule
        & \multicolumn{2}{c}{MAE} & \multicolumn{2}{c}{RMSE} & \multicolumn{1}{c}{MAE Increase (\%)} \\
        \cmidrule(lr){2-3} \cmidrule(lr){4-5} \cmidrule(lr){6-6}
        {Model} & {Normal} & {Crisis} & {Normal} & {Crisis} & {Crisis/Normal-1} \\
        \midrule
        Dynotears           & \textbf{3.39} & \textbf{4.80} & \textbf{4.62} & \textbf{6.64} & 41.59\% \\
        M-Granger           & 3.52 & 4.98 & 4.84 & 6.93 & 41.48\% \\
        SFS-Lin             & 6.22 & 9.74 & 11.97 & 13.61 & 56.75\% \\
        VAR-LiNGAM$^{*}$    & \textbf{3.19} & \textbf{4.52} & \textbf{4.41} & \textbf{6.29} & 41.69\% \\
        \bottomrule
    \end{tabular}
    \caption{Model evaluation metrics ($\times 100$) under normal and crisis market conditions, including the relative increase in MAE.}
    \label{tab:model_metrics}
\end{table}
}

\subsection{Selected features}

\begin{figure}
    \centering
    \includegraphics[width=1\linewidth]{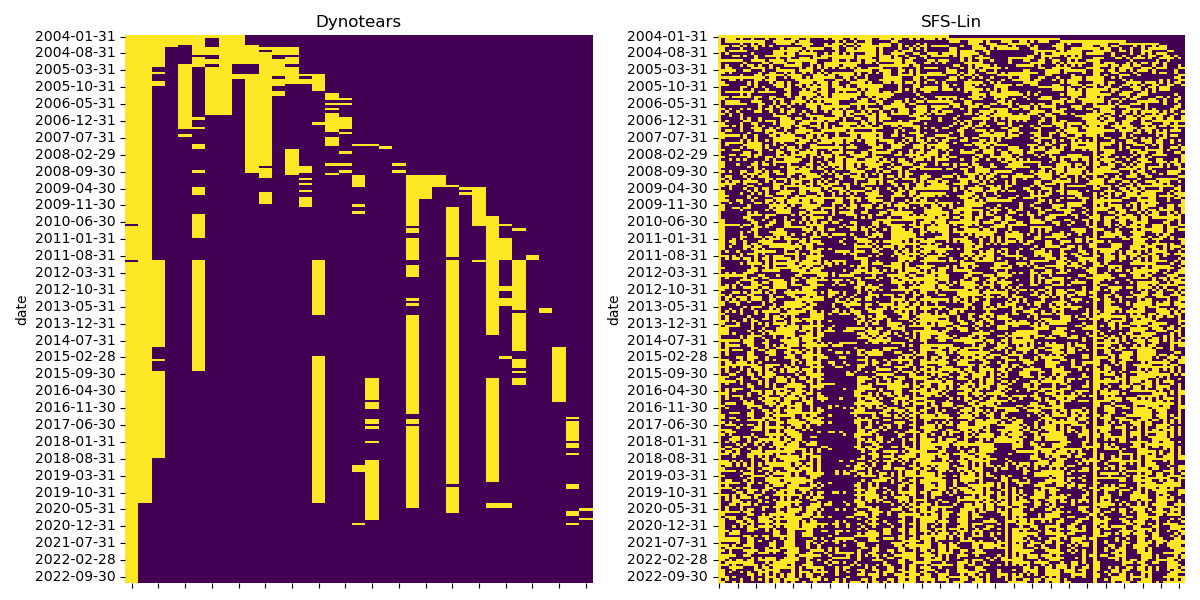}
    \caption{Features selected by Dynotears and SFS-Linear. Yellow pixels indicate selected features, while black pixels indicate unselected features. Note that the number of columns differs between the two plots because only features chosen at least once are included, and many features were never selected by Dynotears throughout the entire period.}
    \label{fig:heat_spy}
\end{figure}

Figure \ref{fig:heat_spy} visualizes the stability of the feature sets selected by Dynotears and SFS-Linear. Without investigating individual variables, the visualization shows that Dynotears selects a relatively consistent set of features from one month to the next, forming a clear pattern. Notably, even during the GFC, the variables selected by Dynotears remain similar to those from the preceding normal period. This consistency is expected because causal parents should persist despite environmental perturbations. 
Starting from the second half of 2019, post-GFC, additional features are selected. This change in the environment provides more information, making previously insignificant causal parents significant due to the increased data and variation. The tests now have adequate power to identify these causal relationships.


On the contrary, the features selected by SFS--Lin are much noisier and lack stability. No features are consistently chosen or eliminated. This suggests that, as a non-causal method, SFS-Linear does not focus on invariance but instead on the strength of correlation, which is subject to random volatility.

We further investigated the features selected by Dynotears leading up to and persisting through the GFC period. The top three features are CPI, personal consumption expenditure, and industrial production: consumer goods. These represent both the demand and supply sides of the economy, along with an indicator for inflation. These indicators are meaningful as they collectively capture the economic conditions impacting market returns, demonstrating the robustness of Dynotears in identifying stable and relevant causal features even during periods of significant economic upheaval. 

\begin{figure}[H]
    \centering
    \includegraphics[width=1\linewidth]{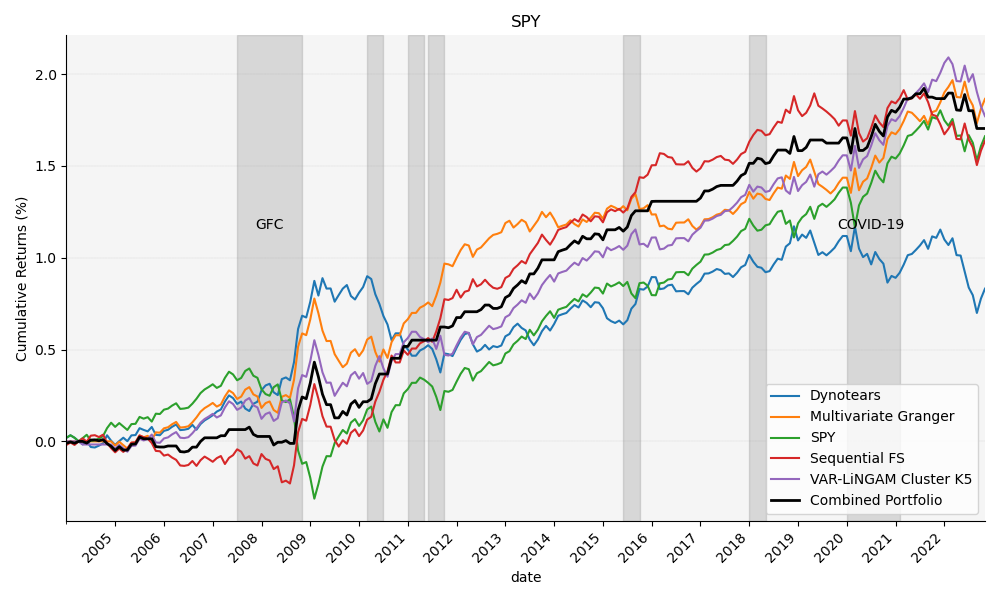}
    \caption{Cumulative returns from trading strategies based on predictions from different feature selection methods. 
    }
    \label{fig:pnl_spy}
\end{figure}
\subsection{Economic value}
Figure \ref{fig:pnl_spy} illustrates the portfolio performance analysis by examining the cumulative returns delivered by each feature selection methodology and a combined portfolio referred to as the ``Combined Portfolio.'' This combined portfolio takes equal weight positions on the SFS--Lin benchmark and the best causal method, the Modified VAR--LiNGAM.

The cumulative return trajectories reveal that the causal methods excel in avoiding large drawdowns for the SPY ETF. For example, Dynotears demonstrated outstanding performance during the Global Financial Crisis (GFC), whereas VAR-LiNGAM maintained relatively impeccable performance during the COVID-19 crisis. Conversely, during periods of market stability, the SFS-Lin performance generally surpasses that of other methods.

Based on the highlighted evidence, it is reasonable to construct a portfolio that balances the trade-off between invariance and prediction, as represented in the causal methods versus the SFS--Lin benchmark. We take equal-weight positions in the SFS--Lin benchmark and the Modified VAR--LiNGAM model to create such a portfolio. The result is depicted in the solid black time series, labeled the ``Combined Portfolio,'' shown in Figure \ref{fig:pnl_spy} and Table \ref{tab:model_metrics_extended}. It can be seen from Table \ref{tab:model_metrics_extended} that the causal methods outperform its benchmark. The M-Granger method is the best method across almost all portfolio metrics. It is interesting to note that, on regular periods, the SFS-Lin metrics are reasonably close, and sometimes even better, as is the case for the Sortino metric, than their causal counterparts. This provides empirical evidence for using a combined portfolio to balance the trade-off we hypothesized earlier. 


{\footnotesize
\begin{table}[h]
    \centering
    \begin{tabular}{@{}lSSSSSS@{}}
        \toprule
        & \multicolumn{2}{c}{E[Rt]} & \multicolumn{2}{c}{Sharpe} & \multicolumn{2}{c}{Sortino} \\
        \cmidrule(lr){2-3} \cmidrule(lr){4-5} \cmidrule(lr){6-7}
        Model & {Normal} & {Crisis} & {Normal} & {Crisis} & {Normal} & {Crisis} \\
        \midrule
        Combined Portfolio  & 9.01 & \textbf{9.86} & \textbf{0.73} & 0.52 & 2.90 & 2.78 \\
        Dynotears           & 4.40 & 8.34 & 0.29 & 0.38 & 1.59 & 2.50 \\
        M-Granger & 9.86 & \textbf{23.44} & 0.67 & \textbf{1.12} & \textbf{3.56} & \textbf{6.75} \\
        SPY                  & 8.78 & -5.56 & 0.59 & -0.25 & 2.52 & -1.12 \\
        SFS-Lin & 8.65 & 10.80 & 0.58 & 0.50 & 3.44 & 3.43 \\
        VAR-LiNGAM$^{*}$  & 9.36 & 8.92 & 0.63 & 0.41 & 3.07 & 2.61 \\
        \bottomrule
    \end{tabular}
    \caption{Model evaluation metrics including E[Rt], Sharpe, and Sortino under normal and crisis periods. The VAR-LiNGAM$^{*}$ represents the modified model with clustering.}
    \label{tab:model_metrics_extended}
\end{table}
}
\subsection{Other causal methods}

We have also explored the performance of seqICP and PCMCI. 

In this application, seqICP consistently returns empty sets, indicating that the algorithm cannot identify causal parents.  This outcome is likely due to the conservative nature of seqICP, as noted by the authors in \citep{pfister2019}, which is designed to be stringent regarding violations of its assumptions. For instance, if there were a direct intervention on the ETF return, such as the GFC directly impacting the ETF return rather than through other variables, the algorithm would reject all sets, resulting in an empty set. This result aligns with the coverage property described in Proposition 2.2 of \citep{pfister2019}, making the output correct but uninformative.

We have also implemented the PCMCI method, which faces the challenge of computational intensity. As noted in \cite{runge2019}, the PC component is significant, leading to the proposal of a faster algorithm that only tests the few covariates with the strongest correlation in the first stage. However, this remains a practical challenge. Running PCMCI on SPY over approximately 200-time steps for 136 predictors takes approximately two weeks to complete. 

\section{Conclusion} \label{sec:conclusion}

In this paper, we develop a pipeline to forecast financial time series with causal discovery methods for stable feature selections. We make comprehensive comparisons of state-of-the-art causal discovery algorithms. Our empirical study confirms that identifying causal features is superior to traditional feature selection methods in return prediction, especially during tumultuous periods, such as financial crises. We have found empirical evidence that causal feature selection methods have been more stable regarding prediction error in most financial crises we labeled. The prediction error metrics that we compare show lower values for all the causal feature selection models when compared to the prediction-based features selection model. Finally, we have found some evidence that, indeed, it seems that a trade-off between invariant-based and prediction-based feature selection models does exist. This is expressed when we compare the portfolio metrics during regular and crisis periods. We found that even though the performance of the SFS-Lin benchmark is close, and for some metrics even better than its causal counterparts, during regular periods, it decreases significantly  during crisis periods.

\bibliographystyle{abbrvnat}
\bibliography{bibliography}

\begin{thebibliography}{57}
\providecommand{\natexlab}[1]{#1}
\providecommand{\url}[1]{\texttt{#1}}
\expandafter\ifx\csname urlstyle\endcsname\relax
  \providecommand{\doi}[1]{doi: #1}\else
  \providecommand{\doi}{doi: \begingroup \urlstyle{rm}\Url}\fi

\bibitem[Ang and Bekaert(2007)]{ang2007stock}
A.~Ang and G.~Bekaert.
\newblock Stock return predictability: Is it there?
\newblock \emph{The Review of Financial Studies}, 20\penalty0 (3):\penalty0 651--707, 2007.

\bibitem[Ariyo et~al.(2014)Ariyo, Adewumi, and Ayo]{ariyo2014stock}
A.~A. Ariyo, A.~O. Adewumi, and C.~K. Ayo.
\newblock Stock price prediction using the arima model.
\newblock In \emph{2014 UKSim-AMSS 16th international conference on computer modelling and simulation}, pages 106--112. IEEE, 2014.

\bibitem[Asness et~al.(2000)Asness, Porter, and Stevens]{asness2000predicting}
C.~S. Asness, R.~B. Porter, and R.~L. Stevens.
\newblock Predicting stock returns using industry-relative firm characteristics.
\newblock \emph{Available at SSRN 213872}, 2000.

\bibitem[{\c{C}}akmakl{\i} and van Dijk(2016)]{ccakmakli2016getting}
C.~{\c{C}}akmakl{\i} and D.~van Dijk.
\newblock Getting the most out of macroeconomic information for predicting excess stock returns.
\newblock \emph{International Journal of Forecasting}, 32\penalty0 (3):\penalty0 650--668, 2016.

\bibitem[Campbell and Shiller(1988)]{campbell1988dividend}
J.~Y. Campbell and R.~J. Shiller.
\newblock The dividend-price ratio and expectations of future dividends and discount factors.
\newblock \emph{The Review of Financial Studies}, 1\penalty0 (3):\penalty0 195--228, 1988.

\bibitem[Campbell and Thompson(2008)]{campbell2008predicting}
J.~Y. Campbell and S.~B. Thompson.
\newblock Predicting excess stock returns out of sample: Can anything beat the historical average?
\newblock \emph{The Review of Financial Studies}, 21\penalty0 (4):\penalty0 1509--1531, 2008.

\bibitem[Campbell and Yogo(2006)]{campbell2006efficient}
J.~Y. Campbell and M.~Yogo.
\newblock Efficient tests of stock return predictability.
\newblock \emph{Journal of financial economics}, 81\penalty0 (1):\penalty0 27--60, 2006.

\bibitem[Chinco et~al.(2019)Chinco, Clark-Joseph, and Ye]{chinco2019sparse}
A.~Chinco, A.~D. Clark-Joseph, and M.~Ye.
\newblock Sparse signals in the cross-section of returns.
\newblock \emph{The Journal of Finance}, 74\penalty0 (1):\penalty0 449--492, 2019.

\bibitem[Chordia and Subrahmanyam(2004)]{chordia2004order}
T.~Chordia and A.~Subrahmanyam.
\newblock Order imbalance and individual stock returns: Theory and evidence.
\newblock \emph{Journal of Financial Economics}, 72\penalty0 (3):\penalty0 485--518, 2004.

\bibitem[Chordia et~al.(2002)Chordia, Roll, and Subrahmanyam]{chordia2002order}
T.~Chordia, R.~Roll, and A.~Subrahmanyam.
\newblock Order imbalance, liquidity, and market returns.
\newblock \emph{Journal of Financial economics}, 65\penalty0 (1):\penalty0 111--130, 2002.

\bibitem[Colombo et~al.(2014)Colombo, Maathuis, et~al.]{colombo2014order}
D.~Colombo, M.~H. Maathuis, et~al.
\newblock Order-independent constraint-based causal structure learning.
\newblock \emph{J. Mach. Learn. Res.}, 15\penalty0 (1):\penalty0 3741--3782, 2014.

\bibitem[Cont(2001)]{cont2001empirical}
R.~Cont.
\newblock Empirical properties of asset returns: stylized facts and statistical issues.
\newblock \emph{Quantitative finance}, 1\penalty0 (2):\penalty0 223, 2001.

\bibitem[Cont et~al.(2023)Cont, Cucuringu, and Zhang]{cross_impact_OFI_QF}
R.~Cont, M.~Cucuringu, and C.~Zhang.
\newblock Cross-impact of order flow imbalance in equity markets.
\newblock \emph{Quantitative Finance}, 23\penalty0 (10):\penalty0 1373--1393, 2023.
\newblock \doi{10.1080/14697688.2023.2236159}.

\bibitem[de~Prado(2023)]{de2023causal}
M.~M.~L. de~Prado.
\newblock \emph{Causal Factor Investing: Can Factor Investing Become Scientific?}
\newblock Cambridge University Press, 2023.

\bibitem[Fama and Schwert(1977)]{fama1977asset}
E.~F. Fama and G.~W. Schwert.
\newblock Asset returns and inflation.
\newblock \emph{Journal of financial economics}, 5\penalty0 (2):\penalty0 115--146, 1977.

\bibitem[Flannery and Protopapadakis(2002)]{flannery2002macroeconomic}
M.~J. Flannery and A.~A. Protopapadakis.
\newblock Macroeconomic factors do influence aggregate stock returns.
\newblock \emph{The review of financial studies}, 15\penalty0 (3):\penalty0 751--782, 2002.

\bibitem[Granger(1969)]{granger1969investigating}
C.~W. Granger.
\newblock Investigating causal relations by econometric models and cross-spectral methods.
\newblock \emph{Econometrica: journal of the Econometric Society}, pages 424--438, 1969.

\bibitem[Gu and Kelly(2020)]{gu2020}
S.~Gu and B.~Kelly.
\newblock Empirical asset pricing via machine learning.
\newblock \emph{Review of Financial Studies}, 33\penalty0 (5):\penalty0 2223--2273, 2020.

\bibitem[Hall(1999)]{hall1999correlation}
M.~A. Hall.
\newblock \emph{Correlation-based feature selection for machine learning}.
\newblock PhD thesis, The University of Waikato, 1999.

\bibitem[Henrique et~al.(2019)Henrique, Sobreiro, and Kimura]{henrique2019literature}
B.~M. Henrique, V.~A. Sobreiro, and H.~Kimura.
\newblock Literature review: Machine learning techniques applied to financial market prediction.
\newblock \emph{Expert Systems with Applications}, 124:\penalty0 226--251, 2019.

\bibitem[Hyv{\"a}rinen et~al.(2010)Hyv{\"a}rinen, Zhang, Shimizu, and Hoyer]{hyvarinen2010estimation}
A.~Hyv{\"a}rinen, K.~Zhang, S.~Shimizu, and P.~O. Hoyer.
\newblock Estimation of a structural vector autoregression model using non-gaussianity.
\newblock \emph{Journal of Machine Learning Research}, 11\penalty0 (5), 2010.

\bibitem[Janzing(2019)]{janzig2019}
D.~Janzing.
\newblock Causal regularization.
\newblock In H.~Wallach, H.~Larochelle, A.~Beygelzimer, F.~d\textquotesingle Alch\'{e}-Buc, E.~Fox, and R.~Garnett, editors, \emph{Advances in Neural Information Processing Systems}, volume~32. Curran Associates, Inc., 2019.
\newblock URL \url{https://proceedings.neurips.cc/paper\_files/paper/2019/file/2172fde49301047270b2897085e4319d-Paper.pdf}.

\bibitem[Kothari and Shanken(1997)]{kothari1997book}
S.~P. Kothari and J.~Shanken.
\newblock Book-to-market, dividend yield, and expected market returns: A time-series analysis.
\newblock \emph{Journal of Financial economics}, 44\penalty0 (2):\penalty0 169--203, 1997.

\bibitem[Krauss et~al.(2017)Krauss, Do, and Huck]{krauss2017deep}
C.~Krauss, X.~A. Do, and N.~Huck.
\newblock Deep neural networks, gradient-boosted trees, random forests: Statistical arbitrage on the s\&p 500.
\newblock \emph{European Journal of Operational Research}, 259\penalty0 (2):\penalty0 689--702, 2017.

\bibitem[Kumbure et~al.(2022)Kumbure, Lohrmann, Luukka, and Porras]{kumbure2022machine}
M.~M. Kumbure, C.~Lohrmann, P.~Luukka, and J.~Porras.
\newblock Machine learning techniques and data for stock market forecasting: A literature review.
\newblock \emph{Expert Systems with Applications}, 197:\penalty0 116659, 2022.

\bibitem[Kyono et~al.(2020)Kyono, Zhang, and van~der Schaar]{kyono2020}
T.~Kyono, Y.~Zhang, and M.~van~der Schaar.
\newblock Castle: Regularization via auxiliary causal graph discovery.
\newblock In \emph{Proceedings of the 34th International Conference on Neural Information Processing Systems}, NIPS'20, Red Hook, NY, USA, 2020. Curran Associates Inc.
\newblock ISBN 9781713829546.

\bibitem[Li(2002)]{li2002macroeconomic}
L.~Li.
\newblock Macroeconomic factors and the correlation of stock and bond returns.
\newblock \emph{Available at SSRN 363641}, 2002.

\bibitem[Lin(2018)]{lin2018technical}
Q.~Lin.
\newblock Technical analysis and stock return predictability: An aligned approach.
\newblock \emph{Journal of financial markets}, 38:\penalty0 103--123, 2018.

\bibitem[Lo and MacKinlay(1988)]{lo1988stock}
A.~W. Lo and A.~C. MacKinlay.
\newblock Stock market prices do not follow random walks: Evidence from a simple specification test.
\newblock \emph{The review of financial studies}, 1\penalty0 (1):\penalty0 41--66, 1988.

\bibitem[Lo et~al.(2000)Lo, Mamaysky, and Wang]{lo2000foundations}
A.~W. Lo, H.~Mamaysky, and J.~Wang.
\newblock Foundations of technical analysis: Computational algorithms, statistical inference, and empirical implementation.
\newblock \emph{The journal of finance}, 55\penalty0 (4):\penalty0 1705--1765, 2000.

\bibitem[Lopez-Paz et~al.(2016)Lopez-Paz, Nishihara, Chintala, Scholkopf, and Bottou]{LopezPaz2016}
D.~Lopez-Paz, R.~Nishihara, S.~Chintala, B.~Scholkopf, and L.~Bottou.
\newblock Discovering causal signals in images.
\newblock \emph{2017 IEEE Conference on Computer Vision and Pattern Recognition (CVPR)}, pages 58--66, 2016.
\newblock URL \url{https://api.semanticscholar.org/CorpusID:1847130}.

\bibitem[Lu et~al.(2023)Lu, Reinert, and Cucuringu]{lu2023co}
Y.~Lu, G.~Reinert, and M.~Cucuringu.
\newblock Co-trading networks for modeling dynamic interdependency structures and estimating high-dimensional covariances in us equity markets.
\newblock \emph{arXiv preprint arXiv:2302.09382}, 2023.

\bibitem[Lu et~al.(2024)Lu, Reinert, and Cucuringu]{lu2024trade}
Y.~Lu, G.~Reinert, and M.~Cucuringu.
\newblock Trade co-occurrence, trade flow decomposition and conditional order imbalance in equity markets.
\newblock \emph{Quantitative Finance}, pages 1--31, 2024.

\bibitem[Lütkepohl(2005)]{lutkepohl2005}
H.~Lütkepohl.
\newblock \emph{New Introduction to Multiple Time Series Analysis}.
\newblock Springer, 2005.
\newblock URL \url{https://EconPapers.repec.org/RePEc:spr:sprbok:978-3-540-27752-1}.

\bibitem[Makridakis et~al.(2018)Makridakis, Spiliotis, and Assimakopoulos]{makridakis2018m4}
S.~Makridakis, E.~Spiliotis, and V.~Assimakopoulos.
\newblock The m4 competition: Results, findings, conclusion and way forward.
\newblock \emph{International Journal of Forecasting}, 2018.

\bibitem[McCracken and Ng(2016)]{mccracken2016}
M.~W. McCracken and S.~Ng.
\newblock Fred-md: A monthly database for macroeconomic research.
\newblock \emph{Journal of Business \& Economic Statistics}, 34\penalty0 (4):\penalty0 574--589, Oct 2016.
\newblock Special Issue on Big Data.

\bibitem[Neely et~al.(2014)Neely, Rapach, Tu, and Zhou]{neely2014forecasting}
C.~J. Neely, D.~E. Rapach, J.~Tu, and G.~Zhou.
\newblock Forecasting the equity risk premium: the role of technical indicators.
\newblock \emph{Management science}, 60\penalty0 (7):\penalty0 1772--1791, 2014.

\bibitem[Neftci(1991)]{neftci1991naive}
S.~N. Neftci.
\newblock Naive trading rules in financial markets and wiener-kolmogorov prediction theory: A study of" technical analysis".
\newblock \emph{Journal of Business}, pages 549--571, 1991.

\bibitem[Niklas~Pfister and Peters(2019)]{pfister2019}
P.~B. Niklas~Pfister and J.~Peters.
\newblock Invariant causal prediction for sequential data.
\newblock \emph{Journal of the American Statistical Association}, 114\penalty0 (527):\penalty0 1264--1276, 2019.
\newblock \doi{10.1080/01621459.2018.1491403}.
\newblock URL \url{https://doi.org/10.1080/01621459.2018.1491403}.

\bibitem[Peters et~al.(2016)Peters, B{\"u}hlmann, and Meinshausen]{peters2016}
J.~Peters, P.~B{\"u}hlmann, and N.~Meinshausen.
\newblock Causal inference by using invariant prediction: identification and confidence intervals.
\newblock \emph{Journal of the Royal Statistical Society: Series B (Statistical Methodology)}, 78\penalty0 (5):\penalty0 947--1012, 2016.

\bibitem[Peters et~al.(2017)Peters, Janzing, and Sch{\"o}lkopf]{peters2017}
J.~Peters, D.~Janzing, and B.~Sch{\"o}lkopf.
\newblock \emph{Elements of Causal Inference: Foundations and Learning Algorithms}.
\newblock Adaptive Computation and Machine Learning series. MIT Press, 2017.
\newblock ISBN 9780262037310.
\newblock Hardcover.

\bibitem[Rozeff(1984)]{rozeff1984dividend}
M.~S. Rozeff.
\newblock Dividend yields are equity risk premiums.
\newblock \emph{Journal of Portfolio management}, pages 68--75, 1984.

\bibitem[Runge et~al.(2019{\natexlab{a}})Runge, Bathiany, Bollt, et~al.]{runge2019}
J.~Runge, S.~Bathiany, E.~Bollt, et~al.
\newblock Inferring causation from time series in earth system sciences.
\newblock \emph{Nature Communications}, 10:\penalty0 2553, 2019{\natexlab{a}}.
\newblock \doi{10.1038/s41467-019-10105-3}.

\bibitem[Runge et~al.(2019{\natexlab{b}})Runge, Nowack, Kretschmer, Flaxman, and Sejdinovic]{runge2019detecting}
J.~Runge, P.~Nowack, M.~Kretschmer, S.~Flaxman, and D.~Sejdinovic.
\newblock Detecting and quantifying causal associations in large nonlinear time series datasets.
\newblock \emph{Science advances}, 5\penalty0 (11):\penalty0 eaau4996, 2019{\natexlab{b}}.

\bibitem[Sims(1980)]{sims1980macroeconomics}
C.~A. Sims.
\newblock Macroeconomics and reality.
\newblock \emph{Econometrica: journal of the Econometric Society}, pages 1--48, 1980.

\bibitem[Spirtes and Glymour(1991)]{spirtes1991algorithm}
P.~Spirtes and C.~Glymour.
\newblock An algorithm for fast recovery of sparse causal graphs.
\newblock \emph{Social science computer review}, 9\penalty0 (1):\penalty0 62--72, 1991.

\bibitem[Spirtes and Zhang(2016)]{spirtes-zhang-2016}
P.~Spirtes and K.~Zhang.
\newblock Causal discovery and inference: concepts and recent methodological advances.
\newblock \emph{Applied Informatics}, 3\penalty0 (1):\penalty0 3, 2016.
\newblock \doi{10.1186/s40535-016-0018-x}.
\newblock URL \url{https://doi.org/10.1186/s40535-016-0018-x}.

\bibitem[Stock and Watson(2001)]{stock2001vector}
J.~H. Stock and M.~W. Watson.
\newblock Vector autoregressions.
\newblock \emph{Journal of Economic perspectives}, 15\penalty0 (4):\penalty0 101--115, 2001.

\bibitem[Sulem et~al.(2024)Sulem, Kenlay, Cucuringu, and Dong]{sulem-etal-2024}
D.~Sulem, H.~Kenlay, M.~Cucuringu, and X.~Dong.
\newblock Graph similarity learning for change-point detection in dynamic networks.
\newblock \emph{Machine Learning}, 113\penalty0 (1):\penalty0 1--44, 2024.
\newblock \doi{10.1007/s10994-023-06405-x}.
\newblock URL \url{https://doi.org/10.1007/s10994-023-06405-x}.

\bibitem[Thawornwong and Enke(2004)]{thawornwong2004adaptive}
S.~Thawornwong and D.~Enke.
\newblock The adaptive selection of financial and economic variables for use with artificial neural networks.
\newblock \emph{Neurocomputing}, 56:\penalty0 205--232, 2004.

\bibitem[Tsai and Hsiao(2010)]{tsai2010combining}
C.-F. Tsai and Y.-C. Hsiao.
\newblock Combining multiple feature selection methods for stock prediction: Union, intersection, and multi-intersection approaches.
\newblock \emph{Decision support systems}, 50\penalty0 (1):\penalty0 258--269, 2010.

\bibitem[Tsay(2005)]{tsay2009}
R.~Tsay.
\newblock \emph{Analysis of financial time series}.
\newblock Wiley series in probability and statistics. Wiley-Interscience, Hoboken, NJ, 2. ed. edition, 2005.
\newblock ISBN 978-0-471-69074-0.
\newblock URL \url{http://gso.gbv.de/DB=2.1/CMD?ACT=SRCHA\&SRT=YOP\&IKT=1016\&TRM=ppn+483463442\&sourceid=fbw_bibsonomy}.

\bibitem[Welch and Goyal(2008)]{welch2008comprehensive}
I.~Welch and A.~Goyal.
\newblock A comprehensive look at the empirical performance of equity premium prediction.
\newblock \emph{The Review of Financial Studies}, 21\penalty0 (4):\penalty0 1455--1508, 2008.

\bibitem[Whitney(1971)]{whitney-1971}
A.~W. Whitney.
\newblock A direct method of nonparametric measurement selection.
\newblock \emph{IEEE Transactions on Computers}, 100\penalty0 (9):\penalty0 1100--1103, 1971.

\bibitem[Whittle(1951)]{whittle1951hypothesis}
P.~Whittle.
\newblock Hypothesis testing in time series analysis.
\newblock \emph{(No Title)}, 1951.

\bibitem[Zhang et~al.(2014)Zhang, Hu, Xie, Wang, Ngai, and Liu]{zhang2014causal}
X.~Zhang, Y.~Hu, K.~Xie, S.~Wang, E.~Ngai, and M.~Liu.
\newblock A causal feature selection algorithm for stock prediction modeling.
\newblock \emph{Neurocomputing}, 142:\penalty0 48--59, 2014.

\bibitem[Zhang et~al.(2019)Zhang, Ma, and Wang]{zhang2019forecasting}
Y.~Zhang, F.~Ma, and Y.~Wang.
\newblock Forecasting crude oil prices with a large set of predictors: Can lasso select powerful predictors?
\newblock \emph{Journal of Empirical Finance}, 54:\penalty0 97--117, 2019.

\end{thebibliography}

\end{document}